\documentclass[preprint,prl,aps,singlecolumn,superscriptaddress]{revtex4-2}
\usepackage{bm}
\usepackage[colorlinks=true,linkcolor=blue,citecolor=blue]{hyperref}
\usepackage{times}
\usepackage{amsmath}
\usepackage{amssymb}
\usepackage{amsthm}
\usepackage{amsfonts}
\usepackage{enumerate}
\usepackage{latexsym}
\usepackage{ifpdf}
\usepackage{cancel}
\usepackage{natbib}
\usepackage{psfrag}
\usepackage[utf8]{inputenc}
\usepackage[T1]{fontenc}
\usepackage{mathtools}
\usepackage[thinc]{esdiff}
\newcommand{\beq}{\begin{equation}}
	\newcommand{\eeq}{\end{equation}}
\usepackage{graphicx}
\usepackage{makeidx}
\usepackage{color}
\usepackage{array}
\usepackage{multirow}
\usepackage{siunitx}
\usepackage{physics}

\begin{document}


\title{Ferroelastic domain wall driven glassy dynamics and photoresponse in a SrTiO$_3$ (111) two-dimensional electron gas}

\author{Jyotirmay Maity}
\email{jyotirmaym@iisc.ac.in}
\affiliation{Department of Physics, Indian Institute of Science, Bengaluru 560012, India}
\author{Manav Beniwal}
\affiliation{Department of Physics, Indian Institute of Science, Bengaluru 560012, India}
\author{Pranab Hazra}
\affiliation{Department of Physics, Indian Institute of Science, Bengaluru 560012, India}
\author{Sumit Tomar}
\affiliation{Department of Physics, Indian Institute of Science, Bengaluru 560012, India}
\author{Srimanta Middey}
\email{smiddey@iisc.ac.in}
\affiliation{Department of Physics, Indian Institute of Science, Bengaluru 560012, India}

\begin{abstract}
The (111) orientation of SrTiO$_3$ (STO)-based heterostructures has emerged as a versatile platform for two-dimensional electron gases (2DEGs), yet the influence of STO's polar ferroelastic domain-wall (DW) network on charge transport remains largely unexplored. Here we study AlO$_x$/STO (111) interfaces hosting 2DEGs of different carrier densities. In the lower-density 2DEG, low-temperature transport exhibits thermal hysteresis, slow resistance relaxation, cooling-rate dependence, and memory effects, signaling glass-like electron dynamics. The relaxation is strongly non-monotonic in temperature, increasing below $\sim$80~K, peaking near 25 K, and vanishing below $\sim$12 K, tracking the established evolution of the ferroelastic DW state of bulk STO rather than any electronic energy scale, which identifies the DW network as its origin. These signatures are absent in a second, more metallic interface with much higher carrier density, demonstrating that the electron-DW coupling can be tuned. We further show that polar DWs govern the photoresponse: the persistent photoconductivity rises sharply below $\sim$80 K, coinciding with the onset of DW polarity. Together, these results establish polar ferroelastic DWs as a carrier-density-tunable electronic degree of freedom in STO (111)-based 2DEGs, opening routes to manipulate 2DEG properties through topological defects.
\end{abstract}

\maketitle

\begin{figure} [t!]
	\vspace{-1pt}
	\hspace{0pt}
\centering\includegraphics[width=0.8\linewidth] {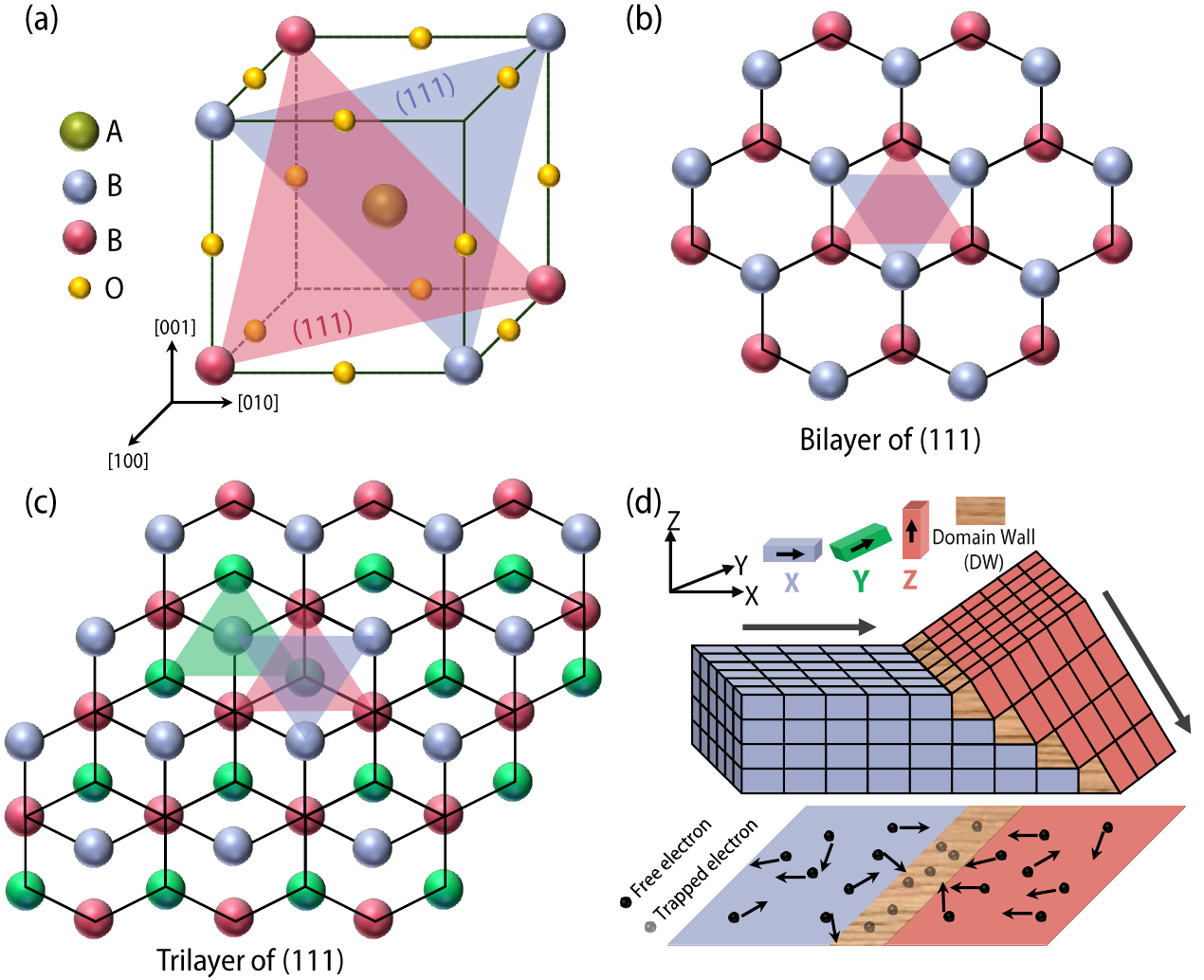}
	\caption{(a) Schematic illustration of the (111) planes formed by the B-site sublattice in an $AB$O$_3$ perovskite structure.  (b), (c) Schematic of bilayer and trilayer of $B$-site lattice geometry along the (111) direction, forming a buckled honeycomb lattice and a dice lattice, respectively. (d) Top panel: schematic of structural twin domains of STO and the domain wall separating them. Three types of domains with the c-axis along [100], [010], and [001] are marked with X (purple), Y (green), and Z (red), respectively. Bottom panel: upon electron doping, itinerant carriers interact with the underlying domain-wall network; free electrons move throughout the domains, while a fraction becomes trapped at the polar domain wall. \label{Fig1} } 
\end{figure}

Geometrical lattice engineering of perovskite oxide ($AB$O$_3$), the growth of heterostructures along low-symmetry directions, provides a powerful route to
achieve emergent quantum phases~\cite{Liu:2016p133}. Along [111], the alternating charged $A$O$_3$ and $B$ planes produce a polar stacking [Fig.~\ref{Fig1}(a)]~\cite{Blok:2011p151917,Middey:2012p261602}.  
Within this
stacking, the $B$-site sublattice forms a buckled honeycomb lattice in the bilayer limit [Fig.~\ref{Fig1}(b)] and a dice lattice in the trilayer [Fig.~\ref{Fig1}(c)], where many emerging phases have been predicted and also demonstrated experimentally~\cite{Xiao:2011p596,Yang:2011p201104,Ruegg:2011p201103,Yang:2011p201104,Doennig:2013p126804,Middey:2016p056801,Kim:2016p7601,Arab:2019p8311}.  In SrTiO$_3$ (STO), whose interfaces with other insulators such as LaAlO$_3$, Al$_2$O$_3$, LaTiO$_3$ etc. famously host a two-dimensional electron gas
(2DEG) together with superconductivity, Rashba spin-orbit coupling,  polar metal behavior~\cite{Ohtomo:2004p423,Reyren:2007p1196,Caviglia:2010p126803, Bert:2011p767,Brehin:2023p823,Cao:2018p1547,Pai:2018p036503}, the (111) orientation has further delivered a correspondingly richer landscape: a sixfold Fermi surface with strongly anisotropic masses and magnetoresistance,  a correlation-driven band inversion, out-of-plane spin textures, electronic nematicity, etc~\cite{Walker:2014p177601,Rout:2017p241107,Monteiro:2019p201102,Khanna:2019p036805,Davis:2017p035127,Davis:2018p041408,Boudjada:2018p086802,He:2018p266802}.

All of these phenomena originate from the interplay among band structure, symmetry and correlations of the confined carriers. The carriers, however, reside within the
STO itself, and this host is far from being a passive substrate: it possesses a structural degree of freedom whose influence on the (111) 2DEG has received far less attention~\cite{Sing:2009p176805,Luca:2018p115143,Schutz:2107p161409}. The cubic-to-tetragonal transition near
105~K in pristine STO produces a dense network of ferroelastic twin domains
separated by domain walls (DWs)~[Fig.~\ref{Fig1}(d)] that become locally polar
below $\sim$80~K while the surrounding domains remain
nonpolar~\cite{Cowley:1964pA981,Fleury:1968p16,Muller:1968p814,shirane:1969p858,Cowley:1969p181,cowley:1996p2799,Scott:2012p187601,Salje:2013p247603}. 
The interplay between this emergent polar order, long-range strain fields, and the intrinsic quantum fluctuations of STO gives rise to a variety of collective DW dynamics at low temperatures~\cite{Pesquera:2018p235701,Kustov:2020p016801,Fauque:2022pL140301}. While metallic carriers are generally expected to screen local dipoles associated with the polar DWs, a growing body of experimental work shows not only that this polarity survives doping but also gives rise to new phenomena through coupling with conduction electrons~\cite{Kalisky:2013p1091,Frenkel:2017p1203,Christensen:2019p269,ojha:2021p054008,Ojha:2026pe74332}. 
Until recently, however, this
coupling was known only through its static consequences such as anomalies in
resistance at the structural transition, carrier trapping, and a zero-field
transverse resistance that is substantially larger in (111)- than in
(001)-oriented heterostructures~\cite{Krantz:2021p036801}. In a recent work on
the metallic $\gamma$-Al$_2$O$_3$/STO (001) 2DEG, we showed that the coupling has also dynamical impact and the slow collective dynamics of the DW network are imprinted directly onto
charge transport, producing  resistance relaxation and memory effects
characteristic of a glassy state~\cite{Maity:2025p84}. This raises two broad questions: how general this glassy behavior is, and whether the
electron-DW coupling can be tuned.

To address these, we investigate AlO$_\text{x}$/STO (111) interfaces hosting 2DEGs with different carrier densities. Our temperature-dependent transport measurements show a clear signature of the underlying ferroelastic DWs in the heterostructure with a room temperature carrier concentration ($n_s$) of 2.3 X 10$^{14}$ cm$^{-2}$. At low temperatures, the resistance exhibits thermal hysteresis and slow temporal relaxation along with cooling-rate dependence and memory effects, consistent with the emergence of a glass-like state below $\sim$40 K~\cite{Amir:2011p235,Pollak:2013p,Kustov:2020p016801}. Interestingly, these signatures are completely absent in a heterostructure with a room temperature $n_s$ of 1.55 X 10$^{16}$ cm$^{-2}$. We further demonstrate that polar DWs have a pronounced influence on the photoresponse of the 2DEG. The system exhibits a large change in photoconductivity that increases sharply below $\sim$80 K, coinciding with the onset of polarity at STO DWs. 
Furthermore, the persistent photoresistance consists of two distinct dynamical contributions, one from midgap states associated with oxygen-vacancy (OV) clustering, the other from the polar DWs, with the DW-related contribution enhanced sharply below the same $\sim$80~K. Together, these results establish that polar ferroelastic DWs act as a carrier-density-tunable electronic degree of freedom in STO~(111)-based 2DEGs.

\begin{figure*} [t!]
	\vspace{-1pt}
	\hspace{0pt}
\centering\includegraphics[width=1\linewidth] {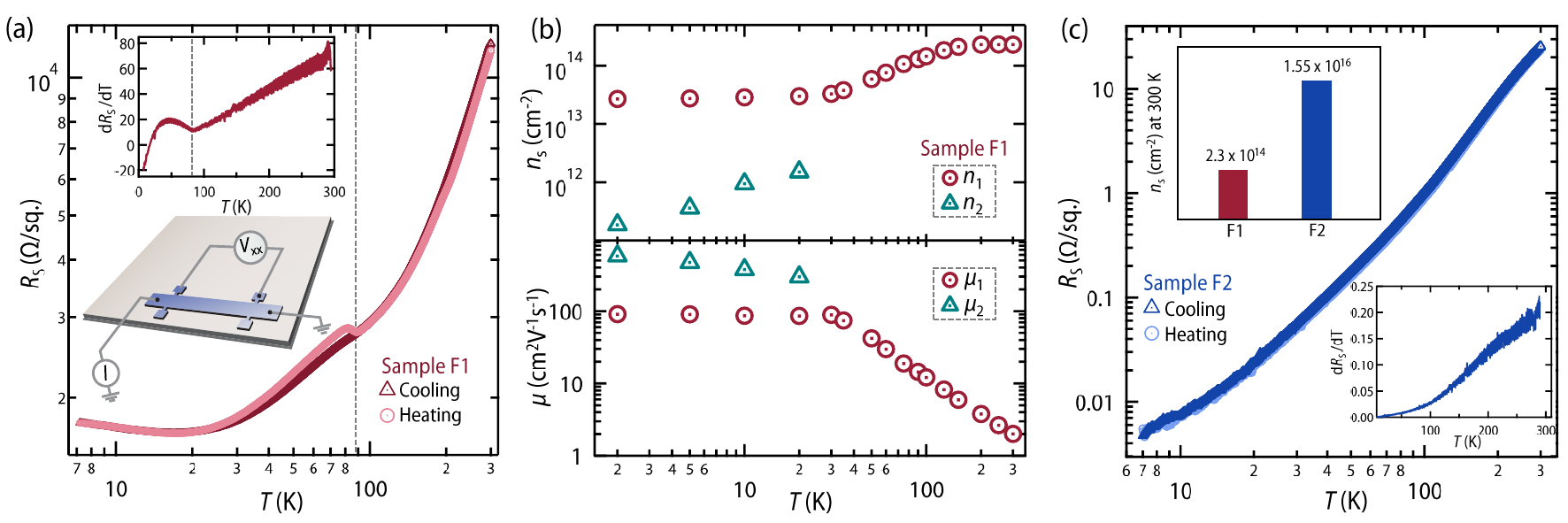}
	\caption{(a) Temperature-dependent sheet resistance ($R_\text{S}$) of the AlO$_\text{x}$/STO (111) heterostructure (F1) measured during cooling and heating cycles. The inset shows the derivative of $R_\text{S}$ with respect to $T$ in the cooling cycle and a schematic of the Hall-bar device geometry used for transport measurements. (b) Temperature-dependent sheet carrier density (n$_\text{s}$, top panel) and mobility ($\mu$, bottom panel) for sample F1, obtained from Hall measurements.
    (c). Temperature-dependent $R_\text{S}$ of another AlO$_\text{x}$/STO (111) heterostructure (F2) measured during cooling and heating cycles. Bottom inset shows the temperature derivative of $R_\text{S}$ ($dR_\text{S}$/dT) during the cooling cycle for sample F2. Top inset shows a comparison of carrier density at 300 K for samples F1 and F2. \label{Fig2}} 
\end{figure*}

AlO$_\text{x}$/SrTiO$_3$ (111) heterostructures were fabricated by ablating a single-crystalline Al$_2$O$_3$ target using a pulsed laser deposition system on mixed-terminated (111)-oriented SrTiO$_3$ substrates ($5 \times 5 \times 0.5\,\text{mm}^3$), commercially available from Shinkosha Co., Ltd., Japan. The deposition was carried out under a vacuum of $\sim$$10^{-6}$ Torr, at a substrate temperature of $500^\circ\text{C}$ for sample F1 and $600^\circ\text{C}$ for sample F2. After growth, the samples were cooled to room temperature at a rate of $15^\circ\text{C}$/min under vacuum. Reflection high-energy electron diffraction (RHEED) images confirmed the amorphous nature of the film, and the film thicknesses were determined from lab-based X-ray reflectivity measurements. For all the electrical transport measurements, a Hall bar structure was fabricated by selectively scratching the film down to the STO substrate and ohmic contacts were made using an ultrasonic wire bonder [inset of Fig.~\ref{Fig2}(a)]. All the temperature-dependent resistance relaxation and photoresponse measurements were carried out in an ARS closed-cycle cryostat, while temperature-dependent Hall measurements were performed using a Teslatron PT system (QD Oxford). Electrical measurements were performed in a dc delta-mode configuration using a Keithley 6221 current source and a Keithley 2182A nanovoltmeter. 

The sheet resistance ($R_\text{S}$) vs. temperature plot for the F1 film [Fig.~\ref{Fig2}(a)], measured during both cooling and heating cycles (temperature ramp rate: 3 K/minute), confirms the formation of a metallic interface. 
The interfacial carriers originate from the OVs created within the top few layers of STO~\cite{Walker:2014p177601,Krantz:2021p036801,Chen:2013p1371,Maity:2025p84}. While the overall temperature dependence of $R_\text{S}$ is metallic, a noticeable thermal hysteresis appears below around 90 K. Specifically, the heating curve shows a distinct kink near 90 K, which is more evident in the temperature derivative of $R_\text{S}$ of the cooling cycle [inset of Fig.~\ref{Fig2}(a)].
A closely analogous hysteresis for STO~(001)-based interfaces was attributed to the formation of the tetragonal ferroelastic domain structure of STO~\cite{Goble:2017p44361,Maity:2025p84}. 
Its appearance here shows that
the coupling between conducting carriers and the DW network is not confined to a
particular crystallographic orientation, and provides the first indication that the
DW network also influences transport in STO~(111)-based 2DEGs. Fig. \ref{Fig2}(b) shows the sheet carrier density ($n_\text{s}$) and mobility ($\mu$) of the film F1. Above 20 K,  the Hall data are well described by a single-band model, whereas below 20 K a two-band model is required. More importantly, the carrier density of the primary channel ($n_1$) decreases progressively below $\sim$150 K, with the most pronounced reduction occurring below $\sim$100 K, which closely coincides with the
appearance of the STO DWs. The relatively low mobility further indicates that carrier transport is strongly influenced by defect-related scattering, including OVs and polar DWs~\cite{Trier:2018p293002}.

\begin{figure*} [t!]
	\vspace{-1pt}
	\hspace{0pt}
\centering\includegraphics[width=0.9\linewidth] {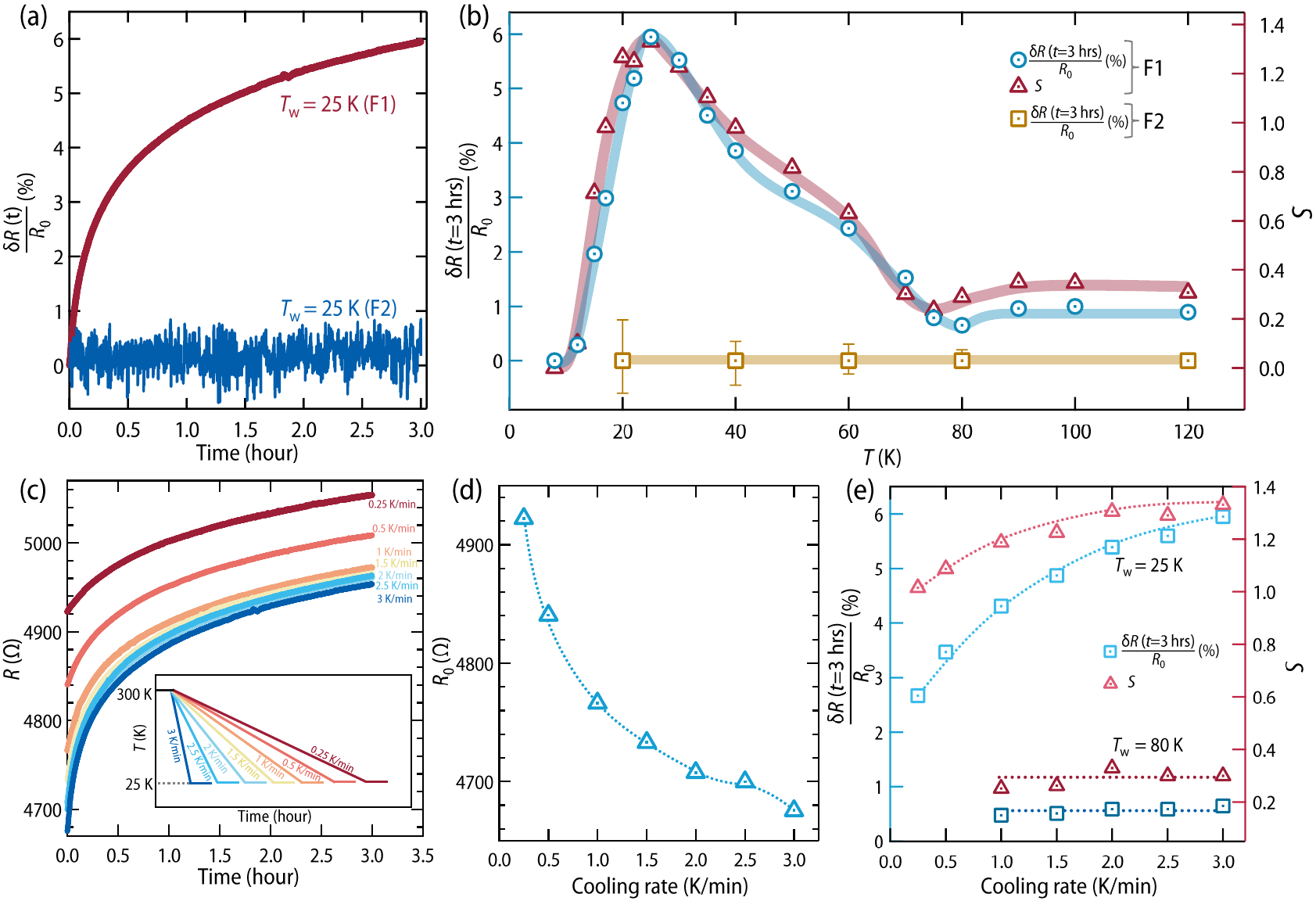}
	\caption{(a) Temporal evolution of resistance at a fixed temperature of 25 K for samples F1 and F2. (b) Temperature dependence of the relative percentage change of resistance after 3 hrs, $\frac{\delta R \textnormal{(t=3 hrs)}}{R_0}(\%)$ (left axis) and relaxation rate,$S$ =$ \frac{d\left[\frac{\delta R(t)}{R_0}\right]}{d\ln(t)}$) (right axis). Here, $\delta R(t)$ = $R (t) - R_0$, where R$_0$ is the resistance at the start of the relaxation. S is obtained by fitting the $\frac{\delta R \textnormal{(t)}}{R_0}(\%)$ vs. $t$ plot in the logarithmic regime. Both quantities exhibit a pronounced maximum near 25 K. $\frac{\delta R \textnormal{(t=3 hrs)}}{R_0}(\%)$ vs. temperature for sample F2 is also shown for comparison. The error bars include the maximum resistance fluctuations observed during the measurement. (c) Temporal evolution of resistance at $T_\text{w}$ = 25 K for different cooling rates. Inset shows the measurement protocol, where we cooled the system from room temperature to 25 K with different cooling rates and waited there for 3 hours to observe the resistance relaxation (d) Initial resistance ($R_0$) as a function of cooling ramp rate, showing a systematic increase in $R_0$ with decreasing cooling rate. (e) Cooling rate dependence of $\frac{\delta R \textnormal{(t=3 hrs)}}{R_0}(\%)$ (left axis) and relaxation rate (S) (right axis) measured at $T_\text{w}$ = 25 K and 80 K for sample F1. A strong dependence of resistance relaxation on cooling rate is observed at 25 K, whereas the effect is negligible at 80 K. \label{Fig3}} 
\end{figure*}

 Having established the static influence of the ferroelastic DW network on charge transport, we next examine whether this coupling can be controlled by the carrier density of the 2DEG. 
 We note that electrostatic gating does not provide a clean carrier-density knob in this system: an out-of-plane field simultaneously restructures the ferroelastic DW network by favoring the nucleation of walls polarized along the field, thereby altering the very configuration whose dynamics we probe. Hence, we grew and investigated another AlO$_x$/STO (111) heterostructure (sample F2) with a much lower sheet resistance and a substantially higher electron concentration [inset and main panel of Fig.~2(c)]. In contrast to F1, F2 shows neither thermal hysteresis nor any anomaly in $dR_S/dT$ [inset of Fig.~2(c)], and the low-temperature carrier freezing is likewise absent. The signatures of the ferroelastic DW network are thus strongly suppressed at higher carrier density; the origin of this suppression is discussed later.

We next focus on probing the dynamical impact of the DWs on the electronic transport as a function of carrier concentration. For this purpose, we cooled down each film from room temperature to a desired fixed temperature ($T_\text{w}$) at a cooling rate of 3 K/min and measured the temporal evolution of resistance. After completing the measurement at a particular $T_\text{w}$, the film was heated to room temperature, before cooling the sample for measurement at the next  $T_\text{w}$.
Fig.~\ref{Fig3}(a) shows the resistance evolution at $T_\text{w} = 25$~K for samples F1 and F2. The resistance of F1 relaxes gradually, exhibiting a logarithmic relaxation at longer time, as commonly observed in conventional electron-glass systems~\cite{Amir:2011p235,Pollak:2013p}. In contrast, F2 shows no measurable relaxation apart from small experimental fluctuations.
To quantify the relaxation, we extract both the relative resistance change after three hours, $\frac{\delta R \textnormal{(t=3 hrs)}}{R_0}(\%)$, and the logarithmic relaxation rate, $S$ =$ \frac{d\left[\frac{\delta R(t)}{R_0}\right]}{d\ln(t)}$). As shown in Fig. \ref{Fig3}(b), both quantities exhibit a similar temperature dependence for F1. The relaxation is weak above $\sim$80 K, increases upon cooling, reaches a maximum near 25 K, and eventually diminishes below 12 K. Interestingly, these characteristic temperatures closely coincide with the established evolution of ferroelastic DW states in bulk STO~\cite{Kustov:2020p016801}. The enhancement below $\sim$80 K follows the onset of polarity at the DWs, while the maximum occurs within the temperature range associated with the domain glass state~\cite{Scott:2012p187601,Salje:2013p247603}. The subsequent suppression below 12 K is consistent with the freezing of DW dynamics in the low-temperature domain solid state~\cite{Kustov:2020p016801}. In contrast, $\frac{\delta R(t=3~\text{hrs})}{R_0}~(\%)$ remains near zero for F2 across the entire temperature range, indicating a complete absence of any glassy relaxation.

 In addition to slow relaxation, glassy dynamics is further characterized by history dependence, whereby the relaxation reflects the path the system has previously taken~\cite{Amir:2011p235,Dobrosavljevic:2012p,Pollak:2013p}. One such measurement is cooling down the system using different cooling rates. Fig.~\ref{Fig3}(c) shows the resulting relaxation curves $R(t)$ for several cooling rates, ranging from 0.25 K/min to 3 K/min for $T_\text{w}$ = 25 K. Two systematic trends are observed. First, the initial resistance $R_0$ increases systematically as the cooling rate decreases [see Fig.~\ref{Fig3}(d)]; second, the subsequent relaxation slows at lower cooling rates. Fig. \ref{Fig3}(e) summarizes the cooling-rate dependence of both $\frac{\delta R \textnormal{(t=3 hrs)}}{R_0}(\%)$ and S for $T_\text{w}$ = 25 K and 80 K for the sample F1. A pronounced dependence on cooling rate is observed at 25 K, which lies within the domain glass state, whereas almost no effect is seen at 80 K, which is outside the glass state. 
 
 In another such history-dependent measurement, we studied the memory effect, where we observed the effect of waiting at a fixed temperature ($T_\text{w}$) before taking a heating cycle. The difference in resistance ($\Delta$) between the wait and no-wait heating cycle shows the memory effect. For $T_\text{w}$ = 25 K,  $\Delta$ exhibits a clear temperature dependence with a dip in d$\Delta$/dT appearing near 25 K, close to the annealing temperature, indicating that the system retains memory of its thermal history. However, for $T_\text{w}$ = 45 K, which lies outside the domain glass state, the memory signal is nearly absent, and no corresponding dip is observed near 45 K. Both the history-dependent measurements thus confirm the presence of a glass-like state at low temperature.

\begin{figure} [htbp]
	\vspace{-1pt}
	\hspace{0pt}
\centering\includegraphics[width=0.85\linewidth] {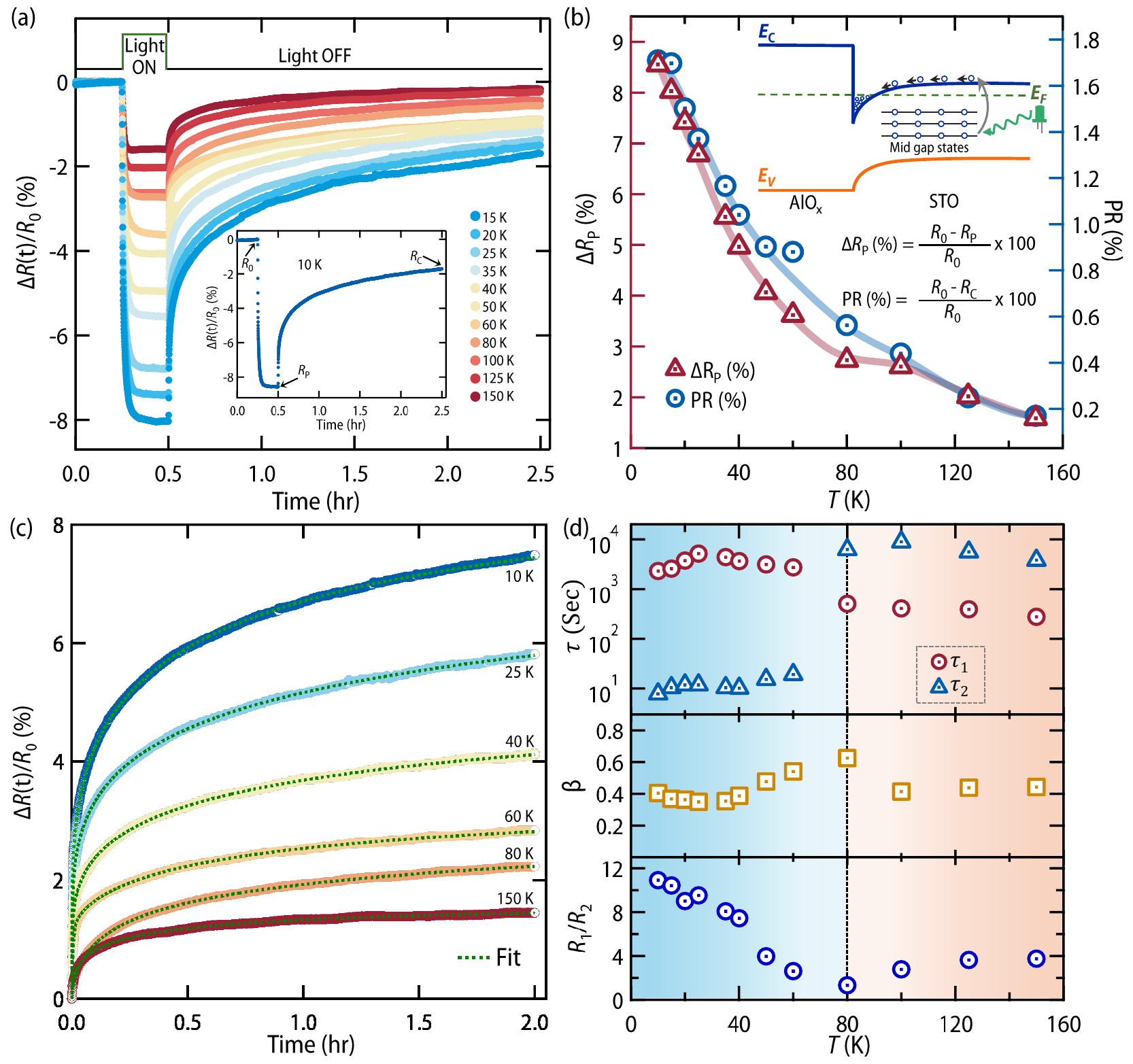}
	\caption{(a) Temporal evolution of resistance at some fixed temperatures under 525 nm green light illumination. The sample was illuminated for 15 min (ON stage), followed by a 2 hr recovery period in the dark (OFF stage). Here, $\Delta R(t)$ is the relative change in resistance. Inset shows a representative measurement at 10 K, indicating the definitions of $R_0$ (equilibrium resistance before light illumination), $R_\text{P}$ (resistance at the end of the light illumination) and $R_\text{C}$ (resistance after 2 hr in darkness). (b)  Photoinduced electron doping ($\Delta R_P(\%)$) ( left axis) and persistent photoinduced resistance change (PR(\%)) (right axis) as a function of temperature. The inset shows a schematic band diagram of the heterostructure illustrating the photoexcitation process from localized mid-gap states into the interfacial conducting states. (c) Normalized change in resistance in the light off-stage as a function of time for different fixed temperatures. Green dotted lines are the fitting curves using equation 1. (d) Temperature dependence of characteristic relaxation times ($\tau_1$ and $\tau_2$), stretching exponent ($\beta$) and relative weight ratio ($R_1$/$R_2$) obtained from the fitting of the recovery curves. All fitting parameters exhibit two distinct temperature-dependent regions separated by a crossover near 80 K. \label{Fig4} } 
\end{figure}

This observation of resistance relaxation and cooling rate dependence can be understood within the framework of a glassy DW state. Once the DWs enter the glassy regime, they no longer equilibrate instantaneously with the cryostat temperature. Consequently, the effective temperature of the DWs ($T_\text{eff}$) always remains higher than the set cryostat temperature ($T_\text{CCR}$)~\cite{Leuzzi:2007thermodynamics,Mauro:2009p75,Leuzzi:2009p686}. As $T_\text{eff}$ gradually approaches thermal equilibrium, the polarity of the DWs slowly increases with time. Since the carrier trapping ability of the DWs is linked to their polarity, gradually more conduction electrons become trapped, leading to the slow increase in resistance as observed during the relaxation measurements. This picture also naturally explains the cooling rate dependence. Slower cooling allows the DWs more time to reach equilibrium. As a result, the DWs possess a larger initial polarization, resulting in stronger carrier trapping and a larger $R_0$. Furthermore, as the system is already closer to equilibrium, the subsequent relaxation is weaker. In contrast, rapid cooling leaves the $T_\text{eff}$ further away from equilibrium, giving rise to a smaller $R_0$ and a larger time-dependent resistance relaxation. Together, the temperature-dependent relaxation, cooling-rate dependence, and the memory effect point to the emergence of a domain glass state between around 12 K and 40 K, while the suppression of the relaxation below 12 K is consistent with the onset of the domain solid state~\cite{Kustov:2020p016801}. The overall temperature dependence of resistance relaxation closely follows the established DW dynamics of bulk STO and is also similar to previous observations in STO~(001)-based 2DEG~\cite{Pesquera:2018p235701,Kustov:2020p016801,Maity:2025p84}.

The results presented above show that the ferroelastic DW network strongly exerts both static and dynamic influences on the transport properties of the F1 sample. Since carrier trapping at the DWs seems crucial to both effects,  we next perturb this trapped population optically.
In STO-based metallic systems, light illumination dopes carriers by exciting electrons from midgap states into the conduction band~\cite{Jin:2015p8778,Niu:2022p2976,Rastogi:2014p125127,Kumar:2015p205117}. These midgap states can arise from OV clusters and from trapping centers at ferroelastic DWs~\cite{Schutz:2107p161409,ojha:2021p054008,Nataf:2020p634}, making photoinduced transport a convenient probe of the interaction between mobile carriers and polar DWs. To study the photoresponse, the sample was illuminated through the cryostat's optical window using a 525 nm green light-emitting diode (LED; purchased from Thorlabs). The optical power incident on the sample was approximately 123 $\si{\micro\watt}$ (measured using a power meter from Coherent: model no. 54-018).  For each measurement, the sample was cooled from room temperature to the target temperature and allowed to stabilize until the resistance drift fell below 0.1\% over a period of 15 min. The sample was illuminated for 15 min, subsequently kept in the dark for a 2-hour recovery period, and its resistance was recorded throughout. The corresponding resistance evolution is shown in Fig. \ref{Fig4}(a). Upon illumination, the initial resistance dropped rapidly and then decayed slowly, indicating photoinduced electron doping of the interfacial 2DEG. Once the light was turned off, resistance tried to get back to its initial state but fell short even after several hours, demonstrating a persistent photoinduced effect~\cite{Jin:2015p8778,Kumar:2015p205117,Gennaro:2013p834}. Such behavior is commonly attributed to the excitations of electrons from localized midgap states into the interfacial quantum well, where spatial separation between photo-excited electrons and remaining trapped holes slows their recombination, leading to long-lived photoconductivity [inset of Fig.~\ref{Fig4}(b)].

To quantify the photoresponse, we define the photoinduced electron doping, $[ \Delta R_P(\%)= (\frac{R_0 - R_{\text{P}}}{R_0}) \times 100]$ and persistent photoinduced resistance change $[\text{PR}(\%)=(\frac{R_0 - R_{\text{C}}}{R_0}) \times 100]$, where $R_0$ is the equilibrium resistance before light illumination, $R_\text{P}$ is the resistance at the end of 15 min light illumination, and $R_\text{C}$ is the resistance measured after two hours in darkness. We plotted the temperature dependence of both quantities in Fig. \ref{Fig4}(b). Both $\Delta R_\text{P}(\%)$ and $\text{PR}(\%)$ increase with decreasing temperature, and exhibit a much stronger enhancement below around 80 K. Notably, this temperature coincides with the onset of polarity at ferroelastic DWs in STO~\cite{Scott:2012p187601,Salje:2013p247603}. While OV-related midgap states are expected to contribute to the photoresponse across all temperatures, the sharp enhancement below 80 K suggests an additional contribution from DW-associated defect states. As the polarity of the DWs develops upon cooling, their ability to localize and trap charge carriers becomes stronger, effectively increasing the density of available photoactive states. Consequently, optical excitation produces a larger carrier modulation and a stronger persistent photoresponse at low temperatures.

The recovery dynamics after turning off the light provide further insight into the microscopic origin of these trapping processes. In many oxide systems, the resistance in light off-stage is well described by a single stretched exponential of the form $R \propto \exp \left[ - \left( \frac{t}{\tau} \right)^\beta \right]$, where  $\beta$ ($0 \le \beta \le 1$) is the stretching exponent and $\tau$ is the relaxation time, respectively~\cite{Rastogi:2014p125127,Kumar:2015p205117}. But a single stretched exponential function is not enough to describe the relaxation over the entire time range. Instead, the recovery resistance is well fitted by a combination of a stretched exponential and a simple exponential component,
\begin{equation}
\resizebox{0.48\textwidth}{!}{$
R(t) = R_0 + R_1 \left[ 1 - \exp \left( -\left( \frac{t}{\tau_1} \right)^\beta \right) \right] + R_2 \left[ 1 - \exp \left( -\frac{t}{\tau_2} \right) \right]$}
\end{equation}
where $R_1$ and $R_2$ represent the amplitudes of the stretched exponential and single exponential components, respectively. Representative fits are shown in Fig. \ref{Fig4}(c). The two relaxation channels naturally suggest two distinct recombination mechanisms. The simple exponential represents recombination through relatively uniform trapping centers with a characteristic relaxation time, consistent with conventional OV-related defect states. In contrast, the stretched exponential component reflects a broad distribution of relaxation times, indicating a spatially heterogeneous energy landscape. Such stretched exponential relaxation is widely regarded as a hallmark of glassy dynamics and is expected when the recombination process is influenced by the evolving polar DW network.

The temperature dependence of $\tau_1$, $\tau_2$ and $\beta$ is shown in Fig. \ref{Fig4}(d). A clear crossover is observed around 80 K. Above this temperature, the two relaxation channels exhibit comparable contributions. However, below 80 K, the stretched exponential component becomes more dominant, as reflected by an increase in the weight ratio $R_1$/$R_2$. Remarkably, this crossover temperature coincides with the onset of polarity at STO DWs. Simultaneously, the characteristic relaxation time $\tau_1$ increases sharply, reaches a maximum near 25 K, and decreases again at lower temperatures, closely mirroring the temperature dependence observed in the resistance-relaxation measurements of Fig. \ref{Fig3}(b).  Since $\tau_1(T)$ and $S(T)$ are extracted from two independent experiments, photoconductivity recovery and dc resistance relaxation, their convergence on the same characteristic temperatures is strong evidence that both are governed by the glassy dynamics of the polar DW network.
In contrast, $\tau_2$ shows only a comparatively weak temperature dependence below 80 K. These observations point to two distinct classes of trapping states: a fast channel ($\tau_2$) from OV defect states present at all temperatures, and a slower glassy channel ($\tau_1$) introduced by polar DWs below the onset of DW polarity. The emergence of this second channel below $\sim$80~K shows that polar DWs directly influence the recombination kinetics of photoexcited carriers in the STO (111)-based 2DEG.

\begin{figure} [t!]
	\vspace{-1pt}
	\hspace{0pt}
\centering\includegraphics[width=0.9\linewidth] {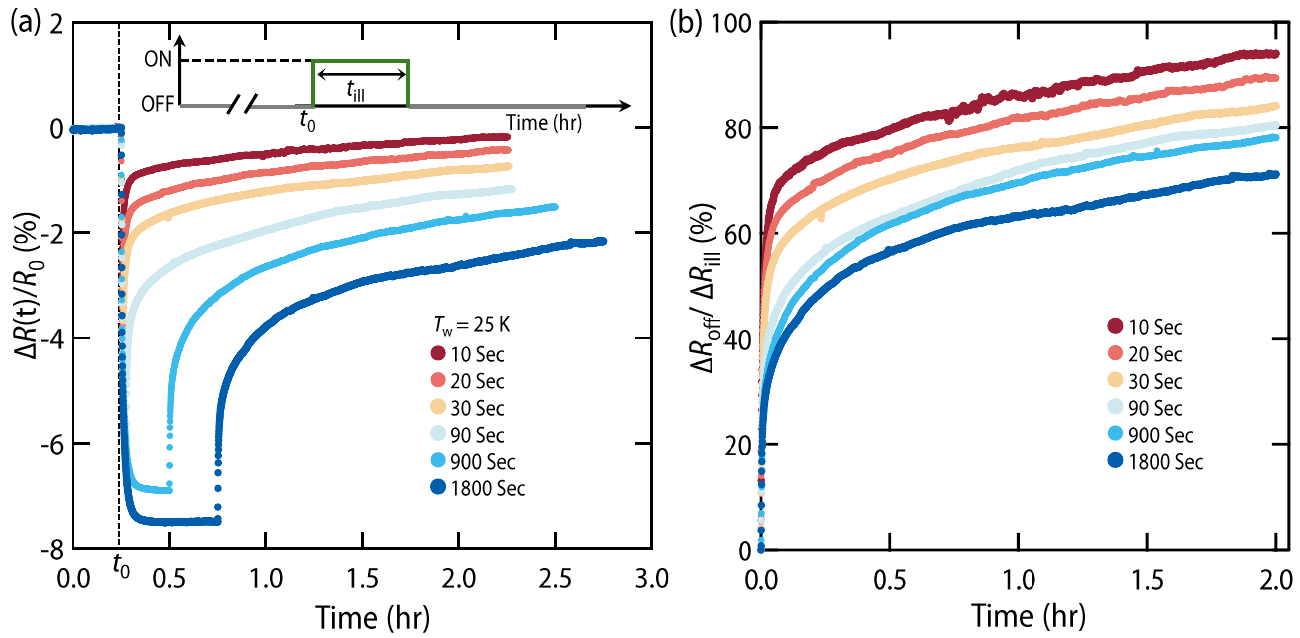}
	\caption{(a) Relative change in resistance $(\frac{\Delta R(t)}{R_0})$ measured at 25 K for different illumination times ($t_\text{ill}$). The inset shows the measurement protocol. The sample was illuminated for a duration $t_\text{ill}$, after which the light was switched off, and the resistance recovery was monitored for 2 hours. (b) Resistance relaxation as a function of time in light off-stage for different $t_\text{ill}$ at 25 K. As different $t_\text{ill}$ produce different photoinduced carrier doping, $\Delta R_\text{off}$ is normalized by the total photoinduced resistance change ($\Delta R_\text{ill}$) after light illumination. \label{Fig5}} 
\end{figure}

Although the slow relaxation, cooling-rate dependence, and memory effect presented above strongly indicate the emergence of a glassy state, another characteristic feature of glassy dynamics is aging, in which relaxation depends on the age of the nonequilibrium state~\cite{Amir:2012p1850,Pollak:2013p}. More specifically, the relaxation behavior of an aged system is slower compared to a younger one as it evolves closer to equilibrium. To investigate whether the photoresponse exhibits such behavior, we performed aging measurements at 25 K, which is well within the domain glass state identified from the resistance relaxation measurements. The measurement protocol is shown in the inset of Fig. \ref{Fig5}(a). The age of the system was controlled by varying the light illumination time ($t_\text{ill}$) from 10 sec to 1800 sec. After each illumination period, the light was turned off, and the resistance recovery was monitored for 2 hours. As shown in Fig. \ref{Fig5}(a), longer illumination times produce a larger photoinduced resistance change, indicating that more carriers are excited from localized states to the conduction band. To compare the relaxation dynamics for different illumination times, we normalize the resistance recovery during the light off-stage ($\Delta R_\text{off}$) by the photoinduced resistance change ($\Delta R_\text{ill}$). The normalized relaxation curves are shown in Fig. \ref{Fig5}(b). Evidently, as the illumination time increases, the system exhibits a slower response, indicating that the response depends on the age of the system, a characteristic signature of aging. 

For systems exhibiting full aging, the relaxation curves are expected to collapse onto a universal curve when plotted as a function of the scaled time $\frac{t}{t_\text{ill}}$~\cite{Amir:2012p1850,Ojha:2024p3830}. However, it is important to note that our system does not entirely follow this simple $\frac{t}{t_\text{ill}}$ scaling. It is worth mentioning that not all glassy systems that exhibit similar aging phenomena strictly follow the $\frac{t}{t_\text{ill}}$ scaling~\cite{Cloitre:2000p4819,Rodriguez:2003p037203}.  Notably, in the present case, the photoresponse contains contributions from both OV-related states and DW-associated trapping states, only the latter exhibiting glassy dynamics. The coexistence of these two relaxation channels likely prevents complete scaling of the data. Moreover, the relative contributions of those two channels are themselves illumination time-dependent, as reflected in the hump-like feature in the relaxation curves whose position shifts with $t_\text{ill}$. This makes it difficult to isolate either component, and further explains why the data do not collapse under simple $\frac{t}{t_\text{ill}}$ scaling. Nevertheless, the systematic slowing down of the recovery with increasing illumination time is consistent with aging and further supports the existence of a glassy DW state in the STO (111)-based 2DEG.

Overall, our work establishes that transport in the STO (111)-based 2DEG is intimately coupled to the polar ferroelastic DW network, and that this coupling is governed by the carrier density: F1 exhibits the full set of signatures-thermal hysteresis, glassy relaxation, cooling-rate dependence, and memory, all of which are strongly suppressed in the more metallic F2 sample. A natural first explanation is that the higher carrier density of F2 enhances the screening of the polar DW fields and weakens their coupling to the itinerant electrons. For a single isotropic band, however, the Thomas-Fermi screening length depends on the effective mass but not on the carrier density, so density alone cannot distinguish the two films within a single-subband picture. In reality, quantum confinement of the STO $t_\text{2g}$ manifold produces a set of discrete subbands, each with distinct orbital character and effective mass~\cite{Khalsa:2013p041302,Walker:2014p177601,Zhong:2013p161102}. As the carrier density increases, $E_\text{F}$ rises and progressively populates additional subbands. Consistent with this picture, our Hall data require a two-band description of F1 at low temperature, confirming multiband occupation. Because the total density of states (DOS) at $E_\text{F}$ is the sum over all occupied subbands and the Thomas-Fermi wavevector $q_\text{TF}$ is set by this total DOS, $q_\text{TF}$ increases each time an additional subband is populated~\cite{Tolsma:2016p045120}. The substantially higher carrier density of F2 is therefore expected to place it beyond such subband thresholds, raising $q_\text{TF}$ and correspondingly shortening the effective screening length relative to F1. Once the screening length falls below the scale of the polar fields surrounding the walls, these fields are screened before they can localize the itinerant electrons, suppressing the DW contribution to transport and accounting for the absence of measurable glassy signatures in F2. 
Additionally, the higher deposition temperature of F2 results in OVs deeper within the substrate rather than only at the interface, consistent with its much higher carrier density~\cite{Herranz:2007p216803,Kalabukhov:2007p121404}. The resulting deeper, quasi-three-dimensional conducting region contributes further in shortening the screening length in F2.

The same polar landscape offers a natural framework for the glassy photoresponse. This parallels our recent work on electron-doped KTaO$_3$, another well-studied quantum paraelectric, where a background dipole glass formed by polar nanoregions likewise renders the electron-hole recombination glassy~\cite{Ojha:2024p3830}. In the present case, illumination excites electrons from localized defect states into the conduction band, and the resulting electron-hole pairs recombine within the slowly evolving fields of the polar DWs. Because electrons and holes are oppositely charged, this background field hinders their recombination and slows their return to equilibrium. As the DW polarity develops below 80~K, the polar landscape becomes increasingly dominant, giving rise to an enhanced persistent photoresponse and the emergence of a stretched-exponential relaxation component. The observation of aging further supports this picture, indicating that the photoexcited carriers inherit the slow dynamics of the underlying glassy DW state.

More broadly, our results add to the growing evidence that ferroelastic DWs act as functional topological defects rather than mere structural imperfections, capable of hosting physics absent from the host lattice itself. A natural next step would be to drive the DW network on purpose using uniaxial strain to deliberately restructure the walls, in combination with optical excitation, and to read out the response electronically, as a direct probe of the interplay between mobile carriers and polar DWs.
Another particularly interesting case is CaTiO$_3$, whose ferroelastic DWs are polar and show antiferroelectric-like switching~\cite{Goncalves:2008p097602,Lu:2025p206103} and which shows OV-driven metallicity with localized in-gap moments~\cite{Mandal:2026p205136}. As the polar walls and structural landscape differ markedly from those of STO, it offers a complementary setting to test whether DW-driven glassy dynamics is a general feature of incipient ferroelectrics upon dilute carrier doping.

	\section*{Acknowledgements}
The authors acknowledge central facilities of the Department of Physics, IISc Bangalore, funded by the FIST program of DST, India. This work is funded by an ANRF ARG grant (grant no. ANRF/ARG/2025/002878/PS) to SM. SM also acknowledges funding from a DST Nano Mission consortium project (No. DST/NM/TUE/QM-5/2019). MB acknowledges funding from the Prime Minister’s Research Fellowship (PMRF), MoE, Government of India. JM acknowledges UGC, India, for the fellowship.  

\section*{DATA AVAILABILITY}
The data that support the findings of this article are not
publicly available. The data are available from the authors
upon reasonable request.

\end{document}